# Three-Point Correlations in the *COBE*[1] DMR Two-Year Anisotropy Maps

G. Hinshaw[2,3], A.J. Banday[2], C.L. Bennett[4], K.M. Górski[5,6], A. Kogut[2]

## ABSTRACT

We compute the three-point temperature correlation function of the *COBE* Differential Microwave Radiometer (DMR) two-year sky maps to search for evidence of non-Gaussian temperature fluctuations. We detect three-point correlations in our sky with a substantially higher signal-to-noise ratio than from the first year data. However, the magnitude of the signal is consistent with the level of cosmic variance expected from Gaussian fluctuations, even when the low order multipole moments, up to $\ell = 9$, are filtered from the data. These results do not strongly constrain most existing models of structure formation, but the absence of intrinsic three-point correlations on large angular scales is an important consistency test for such models.

*Subject headings:* cosmic microwave background — cosmology: observations

## 1. Introduction

The detection of large angular scale anisotropies in the Cosmic Microwave Background (CMB) radiation was first reported by the *COBE*-DMR experiment in 1992 (Smoot et al. 1992; Bennett et al. 1992; Wright et al. 1992; Kogut et al. 1992). The initial detection was based only on the first year of flight data. Since that time the group has processed and analyzed the first two years of data and finds the results to be consistent with the first year results (Bennett et al. 1994). Thus far the analysis of the two-year data has focused primarily on the power spectrum (Górski et al. 1994; Wright et al. 1994a) and its transform, the two-point correlation function (Bennett et al. 1994). In addition to measuring the power spectrum, it is important to determine whether or not the CMB fluctuations are Gaussian. Most of the currently viable theories of structure formation predict that the distribution of CMB temperature fluctuations should appear nearly Gaussian on the large angular scales probed by DMR. For example, standard inflationary models, which generate density fluctuations from quantum fluctuations of an inflation field, predict that CMB temperature fluctuations should be Gaussian to high precision (Guth & Pi 1982; Hawking 1982; Starobinsky 1982; Bardeen, Steinhardt & Turner 1983). On the other hand, models based on topological defects predict deviations from Gaussianity, particularly in the gradient of the temperature field, but generally only on small angular scales ($\lesssim 1°$; Gott et al. 1990; Bennett & Rhie 1993; Perivolaropoulos 1993; Pen, Spergel & Turok 1994; Coulson et al. 1994). Scherrer & Schaefer (1994) show that for a wide class of non-Gaussian density fluctuation fields, the corresponding temperature fluctuations induced by the Sachs-Wolfe effect will appear nearly Gaussian. In light of these expectations, it is especially important to search for non-Gaussian effects in the DMR data: the absence of non-Gaussian features would be an important consistency test for current theories, while the presence of such effects would require a dramatic reformulation of current ideas.

---

[1] The National Aeronautics and Space Administration/Goddard Space Flight Center (NASA/GSFC) is responsible for the design, development, and operation of the Cosmic Background Explorer (*COBE*). Scientific guidance is provided by the *COBE* Science Working Group. GSFC is also responsible for the development of the analysis software and for the production of the mission data sets.

[2] Hughes STX Corporation, Laboratory for Astronomy and Solar Physics, Code 685, NASA/GSFC, Greenbelt MD 20771.

[3] e-mail: hinshaw@stars.gsfc.nasa.gov

[4] Laboratory for Astronomy and Solar Physics, Code 685, NASA/GSFC, Greenbelt MD 20771.

[5] Universities Space Research Assoc., Laboratory for Astronomy and Solar Physics, Code 685.9, NASA/GSFC, Greenbelt MD 20771.

[6] On leave from Warsaw University Observatory, Poland.

Kogut et al. (1995) present an analysis of the two-point correlation function of temperature extrema in the two-year maps. They compare the results to Monte Carlo simulations of Gaussian and non-Gaussian anisotropy models and conclude that, within the class of models tested, the standard Gaussian power law model is the most-likely descriptor of the data. Another readily computable test of Gaussianity is the three-point correlation function, which measures the degree of skewness, or asymmetry in the fluctuations. Numerous authors have recently considered the specific form of the three-point correlation function expected in various cosmological scenarios ranging from standard and non-standard inflationary models to models containing a late-time phase transition (Falk, Rangarajan, & Srednicki 1993; Luo & Schramm 1993; Luo 1994a; Luo 1994b; Gangui et al. 1994; Gangui 1994). All of these authors conclude that the amplitude of such correlations is expected to be small. Thus the presence of non-vanishing three-point correlations in the DMR data would be a clear indication of non-Gaussianity. The three-point correlation function is the average product of three temperatures with a fixed relative separation on the sky: $C_3(\theta_1, \theta_2, \theta_3) = \langle T(\hat{n}_1) T(\hat{n}_2) T(\hat{n}_3) \rangle$ where $\hat{n}_1 \cdot \hat{n}_2 = \cos\theta_1$, $\hat{n}_2 \cdot \hat{n}_3 = \cos\theta_2$ and $\hat{n}_3 \cdot \hat{n}_1 = \cos\theta_3$. Here the angle brackets denote an average over an ensemble of universal observers and, by rotational invariance, this function should depend only on the relative separations $\theta_1$, $\theta_2$, and $\theta_3$. In the absence of access to an ensemble of universal observers, we estimate $C_3$ by performing an average over all directions in our sky. The observed correlation function will thus typically deviate from the universal average we are trying to estimate. We must account for this cosmic variance when deciding if the observed correlation function is evidence for intrinsic three-point correlations in the microwave background. (Thus, by definition, "intrinsic" three-point correlations are only detectable if their amplitude exceeds the level expected from cosmic variance.) Srednicki (1993) has derived the cosmic variance of the three-point function analytically for Gaussian power-law fluctuation models. In this paper, we derive the level of cosmic variance numerically by means of Monte Carlo simulations that account for all important aspects of our data processing such as multipole subtraction on the Galaxy cut sky, and non-uniform pixel weighting. These effects can alter the level of cosmic variance relative to the "ideal" theoretical result, though our calculations agree with Srednicki's in the appropriate limit. Recently Yamamoto & Sasaki (1994) have considered the cosmic variance of the skewness of CMB fluctuations [i.e., $\sigma(C_3(0))$] in a power-law inflation model that generates primordial isocurvature baryon (PIB) fluctuations. They conclude that the mean skewness vanishes and that its cosmic variance is smaller than in adiabatic power-law models by orders of magnitude. Assuming this result applies to the full three-point function, as they suggest, this would imply a level of cosmic variance far less than our current levels of instrument noise. We consider the implications of our current results for this model in the discussion below.

Hinshaw et al. (1994), hereafter Paper I, computed the three-point correlation function of the first-year DMR maps and found evidence for non-vanishing three-point correlations in our sky, albeit with only moderate statistical significance. The amplitude of the correlations was consistent with the level of cosmic variance expected from Gaussian initial fluctuations, and thus could not be deemed intrinsic. In this paper we compute the three-point function of the two-year DMR sky maps. Because of the very significant reduction of noise in the two-year data (by a factor of $\sim 2^{3/2} = 2.8$ relative to the first-year data) we now detect three-point correlations in our sky with high signal-to-noise. However, the amplitude still appears to be consistent with the level of cosmic variance expected from adiabatic, Gaussian power-law models. To minimize this cosmic variance we have applied a series of high-pass filters to the data to remove the low order multipole moments, which are most prone to cosmic variance. In particular we subtract best fit multipole expansions up to order $\ell_{fit} = 9$ to see if the observed signal persists above the diminishing cosmic variance. In most cases the residual signal is statistically significant relative to the instrument noise, but does not exceed the expected level of cosmic variance. Thus we cannot unambiguously interpret the observed three-point correlations as evidence for intrinsic non-Gaussian fluctuations in the CMB.

## 2. Method and Results

The DMR experiment has produced two independent microwave maps (A and B) at each of 3 frequencies (31.5, 53 and 90 GHz). The results presented here are based only on the relatively sensitive 53 and 90 GHz maps. For this analysis we employ an extended Galactic plane cut to further exclude some additional Galactic features that extend beyond the straight 20° cut previously considered, and which could be a source of positive skewness in the data. The extended cut additionally excludes a region at positive latitude near $l = 0°$ (the Ophiuchus complex) and a region at negative latitude near $l = 180°$ (the Orion complex). The resulting map has 3885 remaining pixels. [It is possible to form a significantly less sensitive "reduced Galaxy" map from a suitable linear combination of the three DMR frequency maps (Bennett et al. 1992); however, given the apparent absence of intrinsic three-point correlations in the individual frequency maps, there is little to be gained from analyzing the noisier reduced Galaxy map.] We also subtract a best-fit multipole expansion from the data (see below) before evaluating the three-point function. In §2.1 we compute the equilateral configuration of $C_3$ ($\theta_1 = \theta_2 = \theta_3$), and in §2.2 we compute the "pseudo-collapsed" configuration ($\theta_1 \approx \theta_2, \theta_3 \approx 0$) as described in Paper I. Our notation is as follows: latin indices $i$, $j$, and $k$ refer to sky pixels, while greek indices $\alpha$ and $\beta$ refer to bins of angular separation, of width $2°.6$ degrees.

## 2.1. The Equilateral Case

We define the equilateral three-point correlation function to be

$$C_3^{(e)} = \frac{\sum_{i,j,k} w_i w_j w_k\, T_{1,i} T_{2,j} T_{3,k}}{\sum_{i,j,k} w_i w_j w_k}$$

where the sum is restricted to pixel triples $(i,j,k)$ for which all three pixel separations reside in a single angular separation bin, $T_{1,i}$ is the observed temperature in pixel $i$ of map #1 after multipole subtraction, and so forth, and $w_i$ is the statistical weight of pixel $i$. In the first year data analysis we restricted attention to the auto-correlation functions of the 53 and 90 GHz maps respectively. In the present work we compute both auto and cross-correlation functions, including the 53 GHz auto-correlation, with $T_1 = T_2 = T_3 = 0.5(T_{53A} \pm T_{53B})$, the 53 + 90 GHz auto-correlation, with $T_1 = T_2 = T_3 = 0.5(0.65 T_{53A} + 0.35 T_{90A}) \pm 0.5(0.65 T_{53B} + 0.35 T_{90B})$, and a cross-correlation with $T_1 = T_{53A}$, $T_2 = T_{53B}$, and $T_3 = 0.366 T_{90A} + 0.634 T_{90B}$. (The weights in the 53 + 90 GHz combination were chosen to maximize sensitivity while retaining equal weight in the A and B channels to permit difference map analysis.) With additional integration time, the two-year data are becoming limited by cosmic variance, especially at low multipole order. Thus we adopt uniform pixel weights, $w_i = 1$, to minimize the effects of cosmic variance.

The equilateral three-point functions obtained from the 53 + 90 GHz maps are presented in Figure 1. The data are presented from top to bottom with multipole expansions of increasingly higher order subtracted from the maps before processing. The top panels ($\ell_{min} = 2$) include all moments from quadrupole up, the middle panels ($\ell_{min} = 4$) include all moments from hexadecapole up, while the bottom panels ($\ell_{min} = 10$) include all moments from $\ell = 10$ up. [Fitting a multipole expansion of order $\ell_{fit}$ to data on the cut sky will, upon subtraction, remove some power of order higher than $\ell_{fit}$ because of the loss of spherical harmonic orthogonality (see e.g., Wright et al. 1994b). However, since we apply the same well-defined filter to our fiducial simulations, this effect will be of no practical consequence to our analysis.] The error bars on the data points represent the uncertainty due to instrument noise as determined by 2000 Monte Carlo simulations, while the grey band represents the $rms$ scatter due to a superposition of instrument noise and sky signal, the latter modeled as a scale-invariant ($n = 1$) power law with Gaussian amplitudes and random phases (Bond & Efstathiou 1987), with a mean quadrupole normalization of 20 $\mu$K (Górski et al. 1994). Further details on the simulation methodology may be found in Paper I.

The hypothesis that the computed three-point function is consistent with zero is tested by computing $\chi^2 = \sum_{\alpha,\beta} C_\alpha (M^{-1})_{\alpha\beta} C_\beta$ where $C_\alpha$ is the observed three-point function in angular separation bin $\alpha$, and $M_{\alpha\beta}$ is the covariance matrix computed from the simulated three-point functions. The observed values of $\chi^2$ are given in Table 1, the numbers given in parentheses are the percentage of simulations for which $\chi^2$ exceeded the observed value. Consider first the hypothesis that the observed three-point functions are consistent with $no$ sky signal, ie., that the observed fluctuations are due only to instrument noise. In this case we evaluate $\chi^2$ using the three-point functions observed in the sum maps and the covariance matrix derived from the ensemble of difference maps that have the same noise properties as the sum maps, denoted $\chi^2_{sn}$ in Table 1. For the most sensitive 53 + 90 GHz auto-correlation, we find $\chi^2_{sn} = 101.6, 87.2$, and 65.9 for the $\ell_{min} = 2, 4$, and 10 data respectively, with 48 degrees of freedom. Typically less than 1% of the difference (i.e., noise) simulations had higher values of $\chi^2$, thus we conclude there exist non-vanishing equilateral three-point correlations in the observed CMB sky at greater than 99% confidence level. As a check on the noise levels in our simulations we have evaluated $\chi^2$ using the three-point functions observed in the difference maps, denoted $\chi^2_{nn}$ in Table 1. We find the $\chi^2$ values to be well within the expected range for all three filters, indicating excellent consistency with the modeled noise (see column 4 of Table 1).

We next test the hypothesis that the fluctuations seen in the three-point function are consistent with the level expected to arise from a superposition of instrument noise and sky signal arising from Gaussian, scale-invariant power law fluctuations. In this case $\chi^2$ is evaluated using the covariance matrix computed from the ensemble of sum maps, denoted $\chi^2_{ss}$; for the 53 + 90 GHz auto-correlation we find $\chi^2_{ss} = 31.4, 45.6$, and 57.7 for the $\ell_{min} = 2, 4$, and 10 data respectively. As indicated in Table 1, these values are within the range seen in the simulations, which we take to indicate that the fluctuations observed in the equilateral three-point function are consistent with a superposition of instrument noise and Gaussian CMB fluctuations. However, it is interesting to note that for all three map combinations considered, $\chi^2_{ss}$ is an increasing function of $\ell_{min}$ when compared to the level of cosmic variance. Indeed, for the 53 GHz auto-correlation with $\ell_{min} = 10$, only 2% of our signal + noise simulations had a higher $\chi^2$ than seen in the data. Moreover the relatively more sensitive 53 + 90 GHz auto-correlation appears to have a form reminiscent of the two-point correlation function, as expected in some inflationary models (see, e.g., Falk et al. 1993). However, the Monte Carlo simulations of the 53 + 90 GHz auto-correlation indicate that this result is not statistically exceptional for Gaussian models. The same may be said of the cross-correlation data (see Table 1), although this combination is not quite sensitive enough to probe the fluctuation levels being considered here. In all likelihood, the high $\chi^2$ value seen in the 53 GHz auto-correlation is a manifestation of a slight excess noise ($\sim 2\sigma$) in the 53 GHz (A+B)/2 map (see, e.g., Figure 1 of Hinshaw, 1995 and Table 2 of Banday et al. 1994). Taken as a whole, we do not feel these results present compelling evidence for intrinsic three-point correlations in the CMB.

It is important to note that these data are still noise dominated at these moderately high multipole orders: compare, for example, the relative size of the error bars (which indicate noise) with the grey band (which indicates signal + noise) in Figure 1. The four-year data will be $\sim 2.8$ times more sensitive than the two-year data, which will permit a significant clarification of the present results.

### 2.2. The Pseudo-Collapsed Case

The pseudo-collapsed three-point function, $C_3^{(pc)}$, is defined as above, but now the sum on $j$ is over all pixels that are nearest neighbors to $i$, and the sum on $k$ is over all pixels (except $j$) within a given angular separation bin of $i$. This nearest-neighbors configuration averages over many more distinct pixel triples than the equilateral configuration does, making it more sensitive than the former with respect to instrument noise.

The pseudo-collapsed three-point functions obtained from the 53 + 90 GHz DMR maps are presented in Figure 2, which has the same format as Figure 1. As with the equilateral case, we test the hypothesis that the data are consistent with vanishing three-point correlations by computing $\chi^2 = \sum_{\alpha,\beta} C_\alpha \left(M^{-1}\right)_{\alpha\beta} C_\beta$ where $C_\alpha$ is now the observed pseudo-collapsed function in angular separation bin $\alpha$, and $M_{\alpha\beta}$ is the covariance matrix computed from the ensemble of simulated, pseudo-collapsed functions. The values of $\chi^2$ are given in Table 2, which has the same format as Table 1. For the 53+90 GHz auto-correlation, the $\chi^2$ values for the hypothesis that the three-point correlations in the sum maps are consistent with instrument noise alone are 820.1, 594.9, and 91.3 for the $\ell_{min} = 2$, 4, and 10 filters, respectively, with 71 degrees of freedom. None of the difference (ie., noise) map simulations, with $\ell_{min} = 2$ or 4, had such large $\chi^2$ values, which reinforces our conclusion from the first-year data analysis that we observe non-vanishing three-point correlations in our sky that cannot be attributed to instrument noise. When we include the effects of cosmic variance, the $\chi^2$ values diminish considerably: we find 47.5, 76.2, and 63.3 for the $\ell_{min} = 2$, 4, and 10 filters, respectively. These values are well within the range seen in the Gaussian simulations indicating that cosmic variance can comfortably explain the observed pseudo-collapsed three-point fluctuations. As with the equilateral configuration, the $\chi^2_{ss}$ values for the 53 GHz auto-correlation increase with increasing $\ell_{min}$, but not so in the 53 + 90 GHz combination. The apparent lack of intrinsic three-point correlations in the relatively sensitive pseudo-collapsed configuration is further evidence that the high $\chi^2$ seen in the 53 GHz equilateral function is probably spurious.

### 3. Conclusions

We have evaluated two configurations of the three-point temperature correlation function for the two year *COBE* DMR sky maps and find evidence for non-zero three-point correlations in the data, even at relatively high multipole order ($\ell \gtrsim 10$). This would appear to rule out the particular model for generating PIB fluctuations considered by Yamamoto & Sasaki (1994). However, we demonstrate that the observed level of fluctuations are consistent with the level expected to result from the superposition of instrument noise and a CMB sky signal arising from a Gaussian, scale-invariant power law model of initial fluctuations, with a quadrupole normalized amplitude of 20 $\mu$K. Given that most conventional models of structure formation predict nearly Gaussian fluctuations on large angular scales, it is important to determine whether non-Gaussian features exist in the DMR data. The fact that we find no evidence for large, intrinsic three-point correlations implies that standard structure formation models pass an important consistency test.

Since the three-point function is a cubic statistic, the noise levels will diminish relatively rapidly ($\propto$ time$^{-3/2}$) with additional data. In the four year 53 GHz maps, the rms noise per 2°.6 angular separation bin will be $\sim (11\text{-}12\ \mu\text{K})^3$ for the equilateral configuration and $\sim (6\text{-}9\ \mu\text{K})^3$ for the pseudo-collapsed configuration, depending on angular separation. The added sensitivity will be particularly important for studying correlations in the regime of relatively high multipole order, which is still dominated by instrument noise in the two-year maps.

We gratefully acknowledge the many people who made this paper possible: the NASA Office of Space Sciences, the *COBE* flight operations team, and all of those who helped process and analyze the data. We thank Charley Lineweaver for his useful comments on an earlier version of this paper. This work was supported in part by NASA Grant #NAS5-32648.

Table 1. $\chi^2$ Values for the Equilateral Configuration[a]

| $\ell_{min}$[b] | $\chi^2_{ss}$ [c] | $\chi^2_{sn}$ [d] | $\chi^2_{nn}$ [e] |
|---|---|---|---|
| | 53 GHz auto-correlation | | |
| 2 | 34.2 (83%) | 74.2 (1.5%) | 44.7 (59%) |
| 4 | 55.5 (23%) | 81.5 (0.3%) | 48.5 (45%) |
| 10 | 71.1 (2%) | 76.8 (0.6%) | 43.9 (63%) |
| | 53A × 53B × 90 GHz cross-correlation | | |
| 2 | 37.5 (81%) | 57.6 (17%) | ⋯ |
| 4 | 43.8 (63%) | 50.2 (38%) | ⋯ |
| 10 | 44.1 (64%) | 47.4 (50%) | ⋯ |
| | 53 + 90 GHz auto-correlation | | |
| 2 | 31.4 (88%) | 101.6 (0%) | 41.7 (70%) |
| 4 | 45.6 (53%) | 87.2 (0.2%) | 40.2 (76%) |
| 10 | 57.7 (18%) | 65.9 (5%) | 42.1 (69%) |

[a]There are 48 bins in the equilateral configuration.

[b]$\ell_{min}$ is the lowest order multipole remaining in the map after subtracting a best-fit multipole of order $\ell_{min} - 1$.

[c]$\chi^2_{ss}$ is the $\chi^2$ computed with the covariance derived from the ensemble of signal + noise maps.

[d]$\chi^2_{sn}$ is the $\chi^2$ computed with the covariance derived from the ensemble of noise maps.

[e]$\chi^2_{nn}$ is the $\chi^2$ computed from the difference map data, (A−B)/2, with the covariance derived from the ensemble of noise maps.

Table 2. $\chi^2$ Values for the Pseudo-collapsed Configuration[a]

| $\ell_{min}$ | $\chi^2_{ss}$ | $\chi^2_{sn}$ | $\chi^2_{nn}$ |
|---|---|---|---|
| | 53 GHz auto-correlation | | |
| 2 | 58.3 (67%) | 464.5 (0%) | 69.6 (49%) |
| 4 | 70.8 (43%) | 290.5 (0%) | 71.2 (43%) |
| 10 | 83.4 (17%) | 109.7 (0.5%) | 61.9 (71%) |
| | 53 + 90 GHz auto-correlation | | |
| 2 | 47.5 (86%) | 820.1 (0%) | 62.2 (69%) |
| 4 | 76.2 (30%) | 594.9 (0%) | 58.0 (80%) |
| 10 | 63.3 (68%) | 91.3 (7%) | 78.4 (25%) |

[a]There are 71 bins in the pseudo-collapsed configuration. See notes to Table 1 for further definitions.

# REFERENCES

Banday, A.J., et al. 1994, ApJ, 436, L99

Bardeen, J., Steinhardt, P., & Turner, M. 1983, Phys.Rev.D, 28, 679

Bennett, C.L., et al. 1992, ApJ, 396, L7

Bennett, C.L., et al. 1994, ApJ, 436, 423

Bennett, D., & Rhie, S.H. 1993, ApJ, 406, L7

Bond, J.R., & Efstathiou, G. 1987, MNRAS, 226, 655

Coulson, D., Ferreira, P., Graham, P., & Turok, N. 1994, Nature, 368, 27

Falk, T., Rangarajan, R., & Srednicki, M. 1993, ApJ, 403, L1

Gangui, A. 1994, Phys.Rev.D, 50, 3684

Gangui, A., Lucchin, F., Matarrese, S., & Mollerach, S. 1994, ApJ, 430, 447

Górski, K.M., Hinshaw, G., Banday, A.J., Bennett, C.L., Wright, E.L., Kogut, A., Smoot, G.F., & Lubin, P. 1994, ApJ, 430, L89

Gott, J.R., Park, C., Juszkiewicz, R., Bies, W.E., Bennett, D. Bouchet, F.R., & Stebbins, A. 1990, ApJ, 352, 1

Guth, A., & Pi, S.-Y. 1982, Phys.Rev.Lett, 49, 110

Hawking, S. 1982, Phys.Lett.B, 115, 295

Hinshaw, G., et al. 1994, ApJ, 431, 1

Hinshaw, G., et al. 1995, in CWRU CMB Workshop: 2 Years After COBE, ed. L. Krauss, World Scientific, in press

Kogut, A., et al. 1992, ApJ, 401, 1

Kogut, A., et al. 1995, ApJ, 439, L29

Luo, X., 1994a, ApJ, 427, L71

Luo, X., 1994b, Phys.Rev.D, 49, 3810

Luo, X., & Schramm, D.N. 1993, Phys.Rev.Lett, 71, 1124

Pen, U.-L., Spergel, D.N., & Turok, N. 1994, Phys.Rev.D, 49, 692

Perivolaropoulos, L. 1993, Phys.Rev.D, 48, 1530

Scherrer, R.J., & Schaefer, R.K. 1994, preprint #OSU-TA-10/94, astro-ph/9407089

Smoot, G.F., et al. 1992, ApJ, 396, L1

Srednicki, M. 1993, ApJ, 416, L1

Starobinsky, A. 1982, Phys.Lett.B, 117, 175

Wright, E.L., et al. 1992, ApJ, 396, L13

Wright, E.L., Smoot, G.F., Bennett, C.L., & Lubin, P.M. 1994a, ApJ, 436, 443

Wright, E.L., Smoot, G.F., Kogut, A., Hinshaw, G., Tenorio, L., Lineweaver, C., Bennett, C.L., & Lubin, P.M. 1994b, ApJ, 420, 1

Yamamoto, K., & Sasaki, M. 1994, ApJ, 435, L83

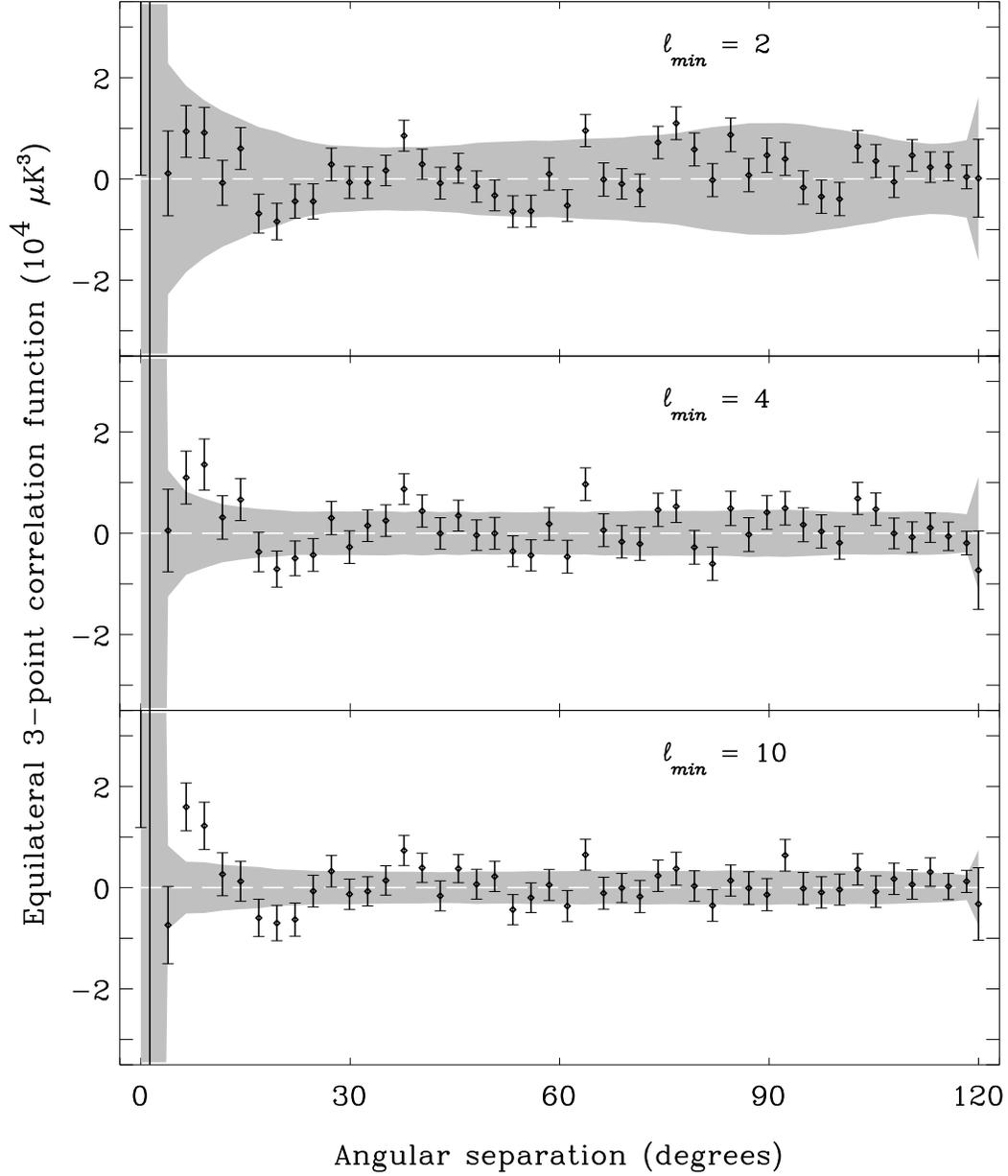

Fig. 1.— *top*) Equilateral three-point correlation function, in thermodynamic temperature units, evaluated from the two-year 53 + 90 GHz average map containing power from the quadrupole moment and up ($\ell_{min} = 2$). The error bars represent the uncertainty due to instrument noise, while the grey band represents the *rms* range of fluctuations expected from a superposition of instrument noise and Gaussian sky signal (see text). *middle*) same as *top* for the map with hexadecapole power and up ($\ell_{min} = 4$). *bottom*) same as *top* for the map with power from $\ell = 10$ and up.

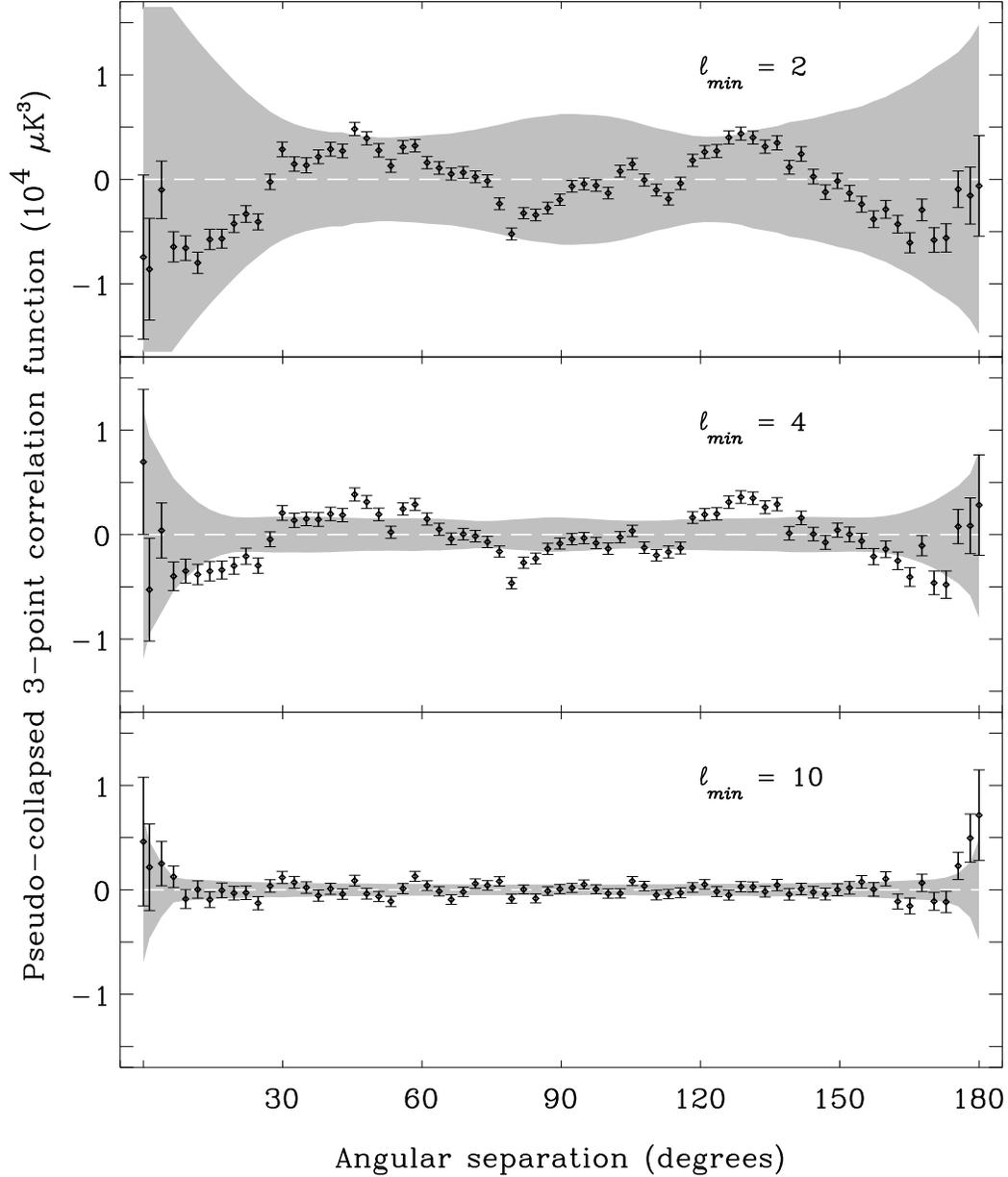

Fig. 2.— *top*) Pseudo-collapsed three-point correlation function, in thermodynamic temperature units, evaluated from the two-year 53 + 90 GHz average map containing power from the quadrupole moment and up ($\ell_{min} = 2$). The error bars represent the uncertainty due to instrument noise, while the grey band represents the *rms* range of fluctuations expected from a superposition of instrument noise and Gaussian sky signal (see text). *middle*) same as *top* for the map with hexadecapole power and up ($\ell_{min} = 4$). *bottom*) same as *top* for the map with power from $\ell = 10$ and up.